\documentclass[%
 reprint,
superscriptaddress,
showpacs,
amsmath,amssymb,
prb,
floatfix,
]{revtex4-1}
\usepackage{graphicx}
\usepackage{subfigure} 
\usepackage[colorlinks,linkcolor=blue,anchorcolor=green,citecolor=blue]{hyperref}
\usepackage{soul}
\usepackage{bm}%

\begin{document}

\title{Positive and negative electrocaloric effect in BaTiO$_3$ in the presence of defect dipoles}

%
%

\author{Yang-Bin Ma}
\affiliation{Institute of Materials Science, Technical University of Darmstadt\ \\ 64287 Darmstadt, Germany}
\author{Anna Gr\"{u}nebohm}
\affiliation{Faculty of Physics and Center for Nanointegration,University of Duisburg-Essen\ \\ 47048 Duisburg, Germany}
\author{Kai-Christian Meyer}
\affiliation{Institute of Materials Science, Technical University of Darmstadt\ \\ 64287 Darmstadt, Germany}
\author{Karsten Albe}
\affiliation{Institute of Materials Science, Technical University of Darmstadt\ \\ 64287 Darmstadt, Germany}
\author{Bai-Xiang Xu}
\affiliation{Institute of Materials Science, Technical University of Darmstadt\ \\ 64287 Darmstadt, Germany}

\date{\today}

\begin{abstract} 
The influence of defect dipoles on the electrocaloric effect (ECE) in acceptor doped BaTiO$_3$ is studied
by means of lattice-based Monte-Carlo simulations. A Ginzburg-Landau type effective Hamiltonian is used. Oxygen vacancy-acceptor associates are described by fixed defect dipoles with
orientation parallel or anti-parallel to the external field.
By a combination of canonical and microcanoncial simulations the ECE is directly evaluated. 
Our results show that in the case of anti-parallel defect dipoles the ECE can be positive or negative depending on the density of defect dipoles. Moreover, a transition from a negative to positive ECE can be observed from a certain density of anti-parallel dipoles on when the external field increases. These transitions are due to the delicate interplay of internal and external fields, and are explained by the domain structure evolution and related field-induced entropy changes.
The results are compared to those obtained by MD simulations employing an {\it{ab initio}} based effective Hamiltonian, and a good qualitative agreement is found.
In addition, a novel electrocaloric cycle, which makes use of the negative ECE and defect dipoles, is proposed to enhance the cooling effect.

\begin{description}
\item[PACS numbers]
77.70.+a, 77.80.-e, 77.80.Jk, 74.62.Dh 
\end{description}
\end{abstract}

\maketitle


\section{Introduction}
Ferroelectric materials based on BaTiO$_3$ 
are of interest for solid state refrigeration because of 
their significant electrocaloric effect (ECE).~\cite{2013_Bai,2013_Moya,2013_Maiwa,2013_Singh,2014_Ye}
Moya {\em et al.}~\cite{2013_Moya} measured the ECE of single crystalline BaTiO$_3$ samples in the direct and indirect way and reported temperature variations of around 1.0\,K.
Qian {\em et al.}~\cite{2014_Qian} investigated the ECE in BaZr$_{0.2}$Ti$_{0.8}$O$_3$ and found a large temperature variation of 4.5\,K over an operating temperature range of 30\,K, which is wider than that in BaTiO$_3$. 
Recently, experimental~\cite{2010_Perantie,2011_Bai,2013_Uddin} and theoretical~\cite{2012_Ponomareva,2013_Li} studies revealed the coexistence of the positive ($\Delta T >0$)
and negative ($\Delta T < 0 $) ECE. 
They concluded that if the external field $\mathbf{E}^{\mathrm{ex}}$ is not parallel with the dielectric polarization of the material, $\mathbf{E}^{\mathrm{ex}}$ may increase the amount of dipolar entropy and thus reduce the phonon contribution under isentropic conditions causing a negative $\Delta T$. 
In first-principles based calculation Ponomareva and Lisenkov~\cite{2012_Ponomareva} found that if the direction and strength of the external field are varied the sign of the ECE can be efficiently controlled near the phase transition temperature. 
Pirc {\em et al.}~\cite{2014_Pirc} studied the negative ECE for an anti-ferroelectric materials based on a generic Kittel model and found transitions from a negative to a positive ECE depending on $\mathbf{E}^{\mathrm{ex}}$ below the Curie temperature. Axelsson {\em et al.}~\cite{2013_Axelsson} modeled the ECE by directly evaluating the entropy from the partition function of a one-dimensional lattice model
and showed how the field applied along a particular lattice direction induces a negative ECE. According to their model the sign reversal of the ECE happens when two phase transition temperatures are close.

Even though the influence of phase transitions on the ECE in ferroelectrics and ferroelectric relaxors was widely studied, there is very few work to reveal the role of dopants and defects on ECE.

In acceptor doped BaTiO$_3$, Ba or Ti ions are substituted by ions with a lower valence. In this case, charge neutrality is typically obtained by compensating oxygen vacancies. Acceptor doped materials, which are also referred to as hard doped, are difficult to polarize, exhibit high coercive fields as well as small strains.~\cite{1971_Jaffe} It was well established that in acceptor doped materials defect reactions between dopants and oxygen vacancies can occur. For BaTiO$_3$ it was demonstrated that ${\rm({Mn''_{Ti}}-V_O^{\bullet \bullet})^\times}$ associates with an excess orientation parallel to the spontaneous polarization are present in reduced crystals.~\cite{1978_Jonker} 
Non-switchable defect dipoles impose a restoring force for reversible domain switching~\cite{2004_Ren} and thus might have an impact on the ECE, since their gradual re-orientation is determined by the barrier for oxygen vacancy migration, hence these defect complexes cannot immediately follow the polarization switching.~\cite{2013_Erhart}
The possibility to enhance and control the ECE in dielectric materials by the presence of internal dipoles was formulated by Van Vechten~\cite{1979_VanVechten} in an US-patent in the late seventies. Only recently, however, Gr{\"u}nebohm {\em et al.}~\cite{2015_Grunebohm} showed by means of Molecular Dynamics (MD) simulations based on an effective Hamiltonian approach that the ECE can switch in the presence of fixed defect dipoles from positive to negative behavior.

In this work, we simulated the influence of the defect dipoles on the ECE, using the MC model presented in our previous work~\cite{2015_Ma}. For the sake of completeness the model is shortly introduced in Sec.~\ref{sec:model}.
The results are presented and discussed in Sec.~\ref{sec:results} with three different subsections.
In Subsec.~\ref{subsec:negative} the influences of different types of defect dipoles on the ECE are investigated.
Results show that the ECE at high temperature can be increased in presence of defect dipoles which lie parallel to the external field (parallel defect dipoles). Moreover, a negative ECE and a double-peak behavior can be induced by defect dipoles which are anti-parallel to the external field (anti-parallel defect dipoles). 
In Subsec.~\ref{subsec:md} we compare the ECE results by MC and MD simulations for different loading scenarios (field-on / field-off). It shows good qualitative agreements on the peculiar ECE behaviors, even though in the MC simulations the strain coupling is absent. 
In Subsec.~\ref{subsec:defect} strategies to tailor the ECE cycle by defect engineering are discussed. 
By applying additionally a reversed electric field during cooling, an enhanced temperature drop is observed in the sample with 3\% defect dipoles.  
Finally, the results are summarized in Sec.~\ref{sec:conclusion}.

\section{Model and Simulation setup} \label{sec:model}

In this study, we analyze the impact of fixed parallel and anti-parallel defect dipoles on the ECE by directly simulating the adiabatic temperature changes using 2-D lattice-based MC simulations based on a Ginzburg-Landau type Hamiltonian for BaTiO$_3$ of our previous work (see Ref.~\onlinecite{2015_Ma}).
The parameters used in this paper are identical to the ones given in Ref.~\onlinecite{2015_Ma}.
For clarification, the model is shortly summarized here.
The total energy of the system $E^*$ involves the thermal energy $E_\mathrm{k}$ and the potential energy $E$:
\begin{eqnarray}
 E^*&=& E_\mathrm{k} + E.
 \label{eq:total_energy}
\end{eqnarray}
The potential energy $E$ is assumed to be of a Ginzburg-Landau type, which includes four contributions:
the Landau multi-well energy term $E_\mathrm{D}$, the dipole-dipole interaction energy
$E_{\mathrm{dip}}$, the domain wall energy $E_{\mathrm{gr}}$ that arises from short-range and elastic interactions,\cite{2008_Nishimatsu} and the electrostatic energy $E_\mathrm{e}$:
\begin{eqnarray}
 E&=& E_\mathrm{D} + E_{\mathrm{dip}}+E_{\mathrm{gr}}+E_\mathrm{e}.
 \label{eq:potential_energy}
\end{eqnarray}
In contrast to the previous work~\cite{2015_Ma}, instead of random fields, defect dipoles are considered.  
The model is extended by introducing fixed non-switchable polarizations on certain lattice sites to mimic the presence of defect dipoles. 
The elastic energy and the electrostrictive interaction are not explicitly included in the Hamiltonian~\cite{2006_Li}, but the elastic energy contribution to domain wall energies is implicitly considered in the gradient term $E_{\mathrm{gr}}$.
However, this simplification has no qualitative influence on the calculated entropy variations, since defect dipoles have only a weak elastic strain field and the possible impact of elastic energies on the domain switching behavior is the same in the isothermal and adiabatic configurations with applied field. 
This statement has been confirmed by a comparison with MD simulations taking the local strain into account (see Subsec.~\ref{subsec:md}).

For the evaluation of the ECE by the MC simulations, there are two methods: the indirect method utilizing Maxwell relations, and the direct method~\cite{2012_Ponomareva} using the modified Creutz's algorithm.
The accuracy of the indirect method depends significantly on the numerical integration.~\cite{2009_Lisenkov}
By contrast, the direct method can sample the change of the temperature accurately, without dependence on the numerical integration.~\cite{2016_Marathe} In this paper the direct method is adopted.

The reason for aging in poled materials is known to origin from the alignment of the dipole defects~\cite{2015_Genenko} and is thus important for possible applications.
However, since the orientation of defect dipoles follows the spontaneous polarization on time scales of days to months depending on temperature, and the simplifying assumption of a fixed dipole orientation within an ECE cycle occurring on much shorter time scales is reasonable. 
Unidirectional defect dipoles might be introduced through a long-time poling process.~\cite{2008_Zhang,2010_Folkman}
Defect dipoles with different orientations are investigated in this paper, namely defect dipoles pointing parallel to $\mathbf{E}^{\mathrm{ex}}$ and defect dipoles oriented anti-parallel to $\mathbf{E}^{\mathrm{ex}}$. 
The local polarization of the defect dipoles~\cite{2011_Eichel} can be approximated as $P_\mathrm{D} = q l / V_0 $, where the charge of the neutral defect dipole in BaTiO$_3$ is $ q = + 2 e $, $l$ is half of the lattice constant $a_0$ and $V_0$ is the lattice volume. Therefore, the local polarization for the defects is assumed as 1.0\,C\,m$^{-2}$, and the direction of these defect dipoles is fixed. 
When several lattice sites are occupied by one type of defect dipole, i.e. the parallel or anti-parallel defect dipoles, the inhomogeneous internal field $\mathbf{E}_\mathrm{i}$ becomes stronger, respectively. Thus,
an impact on the accessible entropy changes can be expected.

It should be noted rather high fields up to 122.2\,kV\,mm$^{-1}$ are applied in this model.
However, extrinsic effects, which is caused by localized nucleation of domains with reversed
polarization in real ferroelectrics, can significantly reduce the coercive field in larger samples or polycrystalline materials.~\cite{2013_Gaynutdinov,2000_Ducharme}

%

Samples with high density of the defect dipoles can be synthesized, e.g. Ba(Ti$_{0.93}$Fe$_{0.07}$)O$_3$~\cite{2012_Lin} and Ba(Ti$_{0.95}$Mn$_{0.05}$)O$_3$~\cite{2011_Shuai} thin films, and the defect density studied in our model ($\leq 6\%$) is within this justifiable region.

%

%

\begin{figure*}[!htp]
 \centering
 \centerline{\includegraphics[width=16cm]{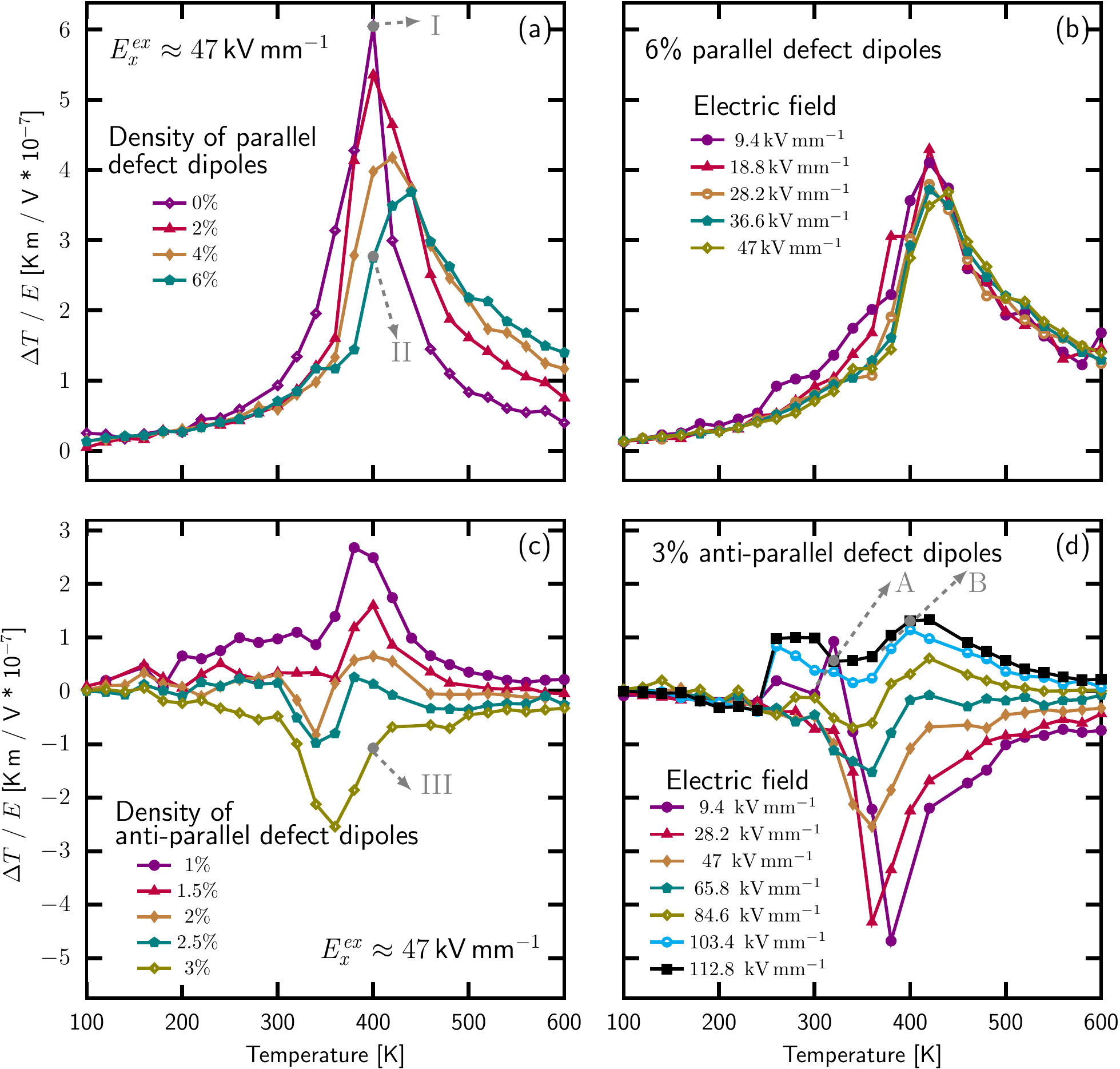}}
 \caption{Positive and negative ECE.
 (a) ECE in the presence of parallel aligned defect dipoles. Different defect concentrations are considered under a given external field $\mathbf{E}^{\mathrm{ex}}$. 
 (b) ECE as a function of $\mathbf{E}^{\mathrm{ex}}$ for a defect concentration of 6\%. 
 (c) Positive and negative ECE in presence of anti-parallel dipoles. When the defect density exceeds a critical value, the resultant internal anti-parallel field overcomes $\mathbf{E}^{\mathrm{ex}}$ and a negative temperature change is observed (negative ECE). 
 (d) Influence of $\mathbf{E}^{\mathrm{ex}}$, while anti-parallel defect dipoles with a concentration of 3\% are present. When $\mathbf{E}^{\mathrm{ex}}$ surpasses the internal field $\mathbf{E}_\mathrm{i}$ induced by the anti-parallel defect dipoles, a negative ECE appears only at lower temperatures, while a positive ECE dominates at higher temperatures, see the olive line (84.6\,kV\,mm$^{-1}$). 
 The above phenomena are explained by the domain structures at the points I, II, III, A and B in Fig.~\ref{fig:dom1}.
 }
 \label{fig:group}
\end{figure*}

\begin{figure*}[!htp]
 \centering
 \centerline{\includegraphics[width=16cm]{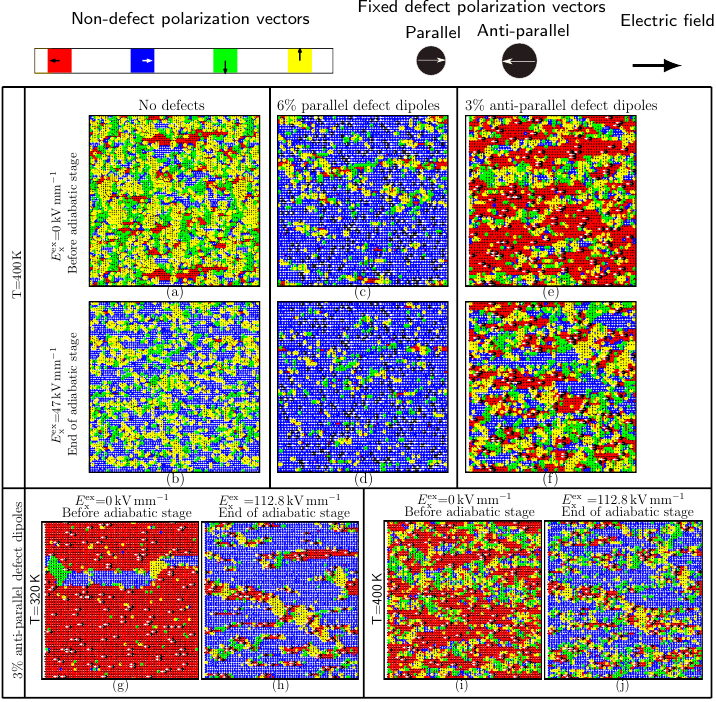}}
 \caption{
 Domain structure explanation of ECE for the points I, II, III, A and B marked in Fig.~\ref{fig:group}. The external field $\mathbf{E}^{\mathrm{ex}}$ is applied in the horizontal direction and points to the right. (a) and (b) are the domain structure snapshots before and after the adiabatic stage for the point I; (c) and (d) for point II; (e) and (f) for point III; (g) and (h) for point A; (i) and (j) for point B. The black dots with white arrows represent the defect dipoles, either parallel or anti-parallel to $\mathbf{E}^{\mathrm{ex}}$. The red, blue, green and yellow dots 
 represent the dipoles pointing respectively to the left, to the right, to the bottom and to the top.}
 \label{fig:dom1}
\end{figure*}

\section{Results and discussion} \label{sec:results}

\subsection{Positive and negative electrocaloric effect} \label{subsec:negative}

For the simulations in this subsection the following loading history is applied.
Firstly, the sample is equilibrated at a given temperature without applying a external field for 28,800 MC steps. 
During one MC step all $L\times L$ sites are visited, and the polarization switching is allowed with a certain probability.
%
%
%
From this equilibrium state the sample is then poled at a fixed external electric field for 28,800 MC steps. 
After instantaneous removal of the poling electric field and a third relaxation step with 28,800 MC steps, the prepoled sample is obtained.
These processes are performed in a canonical ensemble at constant temperature $T_0$. 
Finally, the prepoled sample is used to study the ECE under an adiabatic condition with a field switched on instantaneously.
Hence the microcanonical ensemble with a constant total energy is utilized. After running 43,200 MC steps, the system reaches the equilibrium state with temperature $T_1$, i.e. the end of the adiabatic stage. 
This kind of loading is named the ``Field on" case.
The typical measurement time length of the ECE in experiments is seconds, which is much longer than the characteristic switching time length of the polarization. Therefore, the instantaneous field removal or field application is utilized.
Although the ramping of the external field can give different values for the temperature change, it has been shown that the resultant difference is small. \cite{2016_Marathe} 
In Fig.~\ref{fig:group} the ECE is investigated within the ``Field on" case.

The influence of the concentration of parallel defect dipoles on the ECE is illustrated in Fig.~\ref{fig:group}(a). 
For pure BaTiO$_3$ the magnitude of $\Delta T / \Delta E$ at the peak is $\rm 5.7 \times 10^{-7}$\,K\,m/V, which is comparable with the experimental data $\rm 8.3 \times 10^{-7}$\,K\,m/V by Moya {\em et al.}~\cite{2013_Moya}.
With increasing defect concentration, the ECE peak decreases, and the maximum shifts to higher temperatures, due to the reason that $\mathbf{E}_\mathrm{i}$ induced by the defect dipoles stabilizes the tetragonal phase. However, the interesting feature shown in Fig.~\ref{fig:group}(a) is that the temperature change of the samples with parallel defect dipoles is larger than that in the defect-free one within the high temperature range. 
In addition, the obtainable ECE is enhanced when $\mathbf{E}^{\mathrm{ex}}$ is increased (see Fig.~\ref{fig:group}(b)) since higher $\mathbf{E}^{\mathrm{ex}}$ can induce a more ordered system. 
Similar results have been reported by Rose and Cohen~\cite{2012_Rose}, where no defect dipoles were considered.
It can be concluded that above the phase transition temperature the ECE can be enhanced either by application of higher fields or by incorporation of parallel defect dipoles.

For clarification of the observed effects we study the domain configurations at different points in the cycle.
The prepoled sample shows a multiple domain configuration (Fig.~\ref{fig:dom1}(a)) which is switched 
into a state of reduced dipolar entropy after the field is switched on (Fig.~\ref{fig:dom1}(b)). And thus the temperature is increased under adiabatic conditions by $\mathbf{E}^{\mathrm{ex}}$. 
It can clearly be seen that a smaller number of domains occurs in the presence of $\mathbf{E}^{\mathrm{ex}}$ than in the prepoled sample without $\mathbf{E}^{\mathrm{ex}}$.

If we compare now with the case, where defect dipoles in parallel orientation to $\mathbf{E}^{\mathrm{ex}}$ are present, the situation changes (Fig.~\ref{fig:dom1}(c) and (d)). 
In the initial state the configurational space is reduced as compared to the defect free sample (see Fig.~\ref{fig:dom1}(a) and (c)), since defect dipoles induce $\mathbf{E}_\mathrm{i}$ that locally stabilizes the domain structure against thermal fluctuations. If $\mathbf{E}^{\mathrm{ex}}$ is now switched on, the change in configurational entropy ($S_{\mathrm{conf}}$) is smaller than in the defect-free case and thus a smaller temperature variation is possible, which explains the defect-concentration dependence of the ECE. If fewer domain walls disappear, the decrease of the domain wall energy is reduced and
the excess heat is reduced. 
In contrast to the low temperature range, at higher temperatures the ECE is elevated by the existence of the parallel defects.

A different scenario occurs, if the defect dipoles are pre-aligned in anti-parallel direction (see Fig.~\ref{fig:group}(c)); a situation which can be installed by poling the sample on extended time scales, before reversing
the field direction. For defect concentrations above 2\% the ECE turns from a positive effect into a negative one and becomes more pronounced with increasing $\mathbf{E}^{\mathrm{ex}}$.
This can be explained again from the characteristic domain structure shown in Fig.~\ref{fig:dom1}(e) and (f) with 3\% anti-parallel defect dipoles. Initially, the system is close to a single domain configuration (Fig.~\ref{fig:dom1}(e)) with orientation parallel to the defect dipoles and thus has a small $S_{\mathrm{conf}}$ because of the large average internal field of $\langle \mathbf {E_\mathrm{i}}\rangle=-87.2$\,kV\,mm$^{-1}$.

If now an external field of $\mathbf{E}^{\mathrm{ex}}=47.0$\,kV\,mm$^{-1}$ is switched on, $\mathbf{E}_\mathrm{i}$ is partly compensated and the material can locally sample a multi-domain structure with increased $S_{\mathrm{conf}}$ (see Fig.~\ref{fig:dom1}(f)) corresponding to a reduced potential energy. Thus, vibrational entropy has to be removed from the phonons under isentropic conditions and the
negative ECE appears. 
It should be noted that the domain snapshots, i.e. Fig.~\ref{fig:dom1}(e) and (f), are for 400\,K. At the negative ECE peak, i.e. 360\,K, the prepoled sample is more like a single domain type, and the $S_{\mathrm{conf}}$ change is more prominent.

When the magnitude of $\mathbf{E}^{\mathrm{ex}}$ is, however, comparable with that of $\mathbf{E}_\mathrm{i}$ (hereby between 84.6\,kV\,mm$^{-1}$ and 112.8\,kV\,mm$^{-1}$), both the ECE and the negative ECE start to coexist (see Fig.~\ref{fig:group}(d)). 
A sharp rise in temperature is observed at the transition from the negative into positive temperature change, and both the negative ECE peak and the first positive ECE peak shift to lower temperature with increasing $\mathbf{E}^{\mathrm{ex}}$.

This transition, again, can be explained by inspecting the domain structure for the case of $\mathbf{E}^{\mathrm{ex}}=112.8$\,kV\,mm$^{-1}$. 
Before $\mathbf{E}^{\mathrm{ex}}$ is applied, thermal energy fluctuations provide the ability to sample configurational space in areas, which are not pinned by the presence of local defect dipoles. 
This can be seen in Fig.~\ref{fig:dom1}(g), where the single domain states (red) form a percolating (non switching) network of the sites localized in the direct vicinity of defect dipoles, while multiple domain configurations, which can take various iso-energetic configurations, occur in between. 
If now an $\mathbf{E}^{\mathrm{ex}}$, which is stronger than the internal field, is switched on, the situation is reversing. Sites being part of the initially inactive percolating network around the anti-parallel defect sites (see Fig.~\ref{fig:dom1}(h)) are those where the external field is compensated. Thus, only these sites in the direct vicinity of defect associates can effectively sample configurational space, while the remaining matrix polarizations orient parallel to $\mathbf{E}^{\mathrm{ex}}$ and do not contribute to $S_{\mathrm{conf}}$. Since the relative number of sites which can possibly flip their configuration scales with the defect concentration and is thus smaller than the number of sites dominated by $\mathbf{E}^{\mathrm{ex}}$, $S_{\mathrm{conf}}$ is reduced and the ECE becomes positive again.

Under high enough field (hereby 103.4 and 112.8\,kV\,mm$^{-1}$) and within a low temperature range ($\leq 240$\,K) the negative ECE still persists due to the same reason discussed above for the case under low fields.
When the temperature slightly increases, a sharp transition from this negative to positive ECE is observed as well as in the case with 84.6\,kV\,mm$^{-1}$.
However, with further increasing of the initial temperature, $\Delta T/E$ in Fig.~\ref{fig:group}(d) shows a cusp (around 320\,K under 112.8\,kV\,mm$^{-1}$) and a hump (around 400\,K under 112.8\,kV\,mm$^{-1}$).
This double-peak behavior can be attributed to a temperature induced phase transition in the initial state. 
As can be seen from the domain configuration, the local domains have formed at $T=400$\,K and the system has locally taken a multi-domain configuration in the initial prepoled sample, which is higher in potential energy than the single domain case prevailing below $T=320$\,K. Thus the effective $\Delta T$ increases again. 
In brevity, with increasing initial temperature this local order is further reduced and the material behaves like a regular ferroelectric material.

\subsection{Comparison between MC and MD} \label{subsec:md}

\begin{figure*}[htp]
 \centering
 \centerline{\includegraphics[width=16cm]{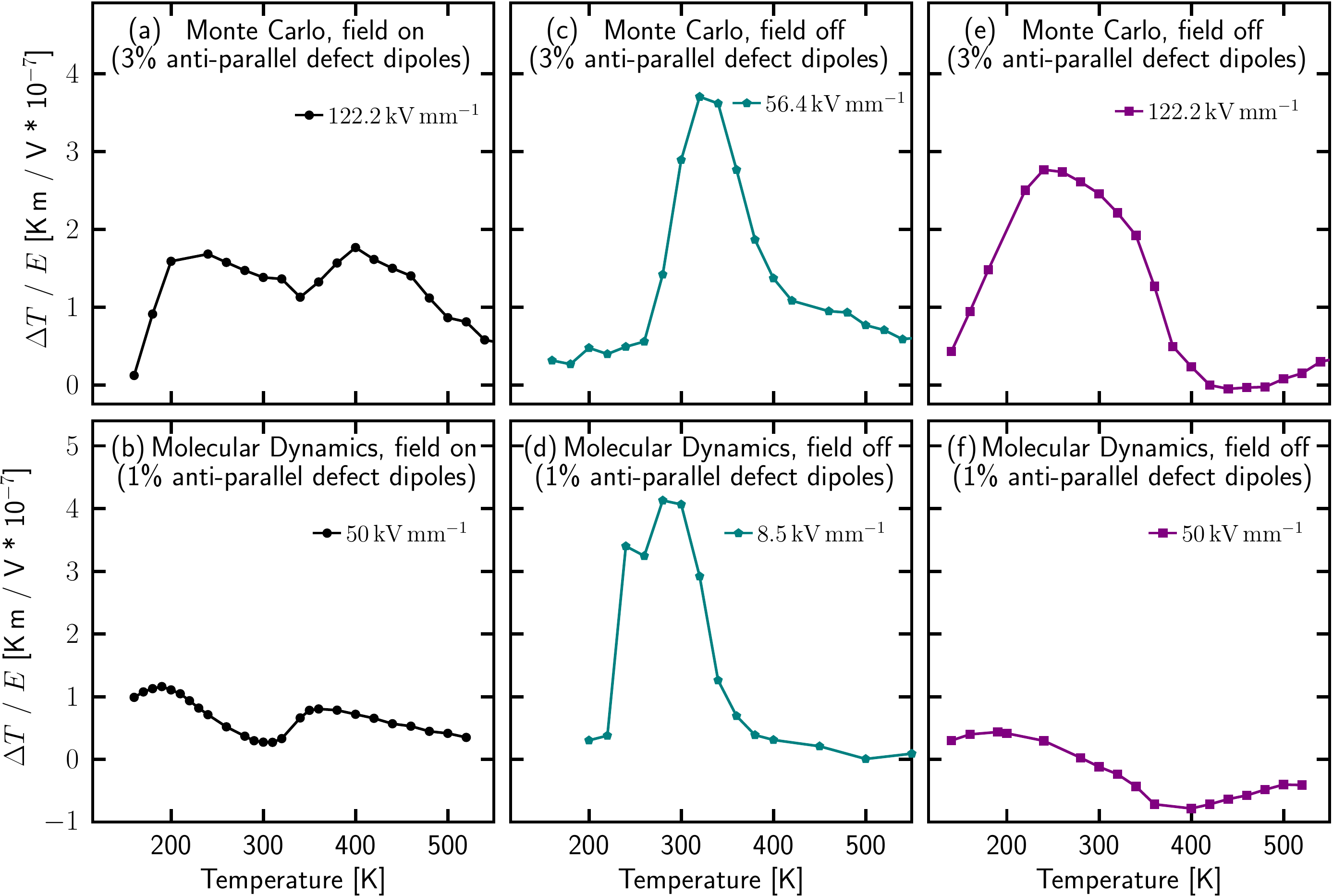}}
 \caption{
 Comparison of $\Delta T /E$ over $T$ from the MC simulations with 3\% anti-parallel defect dipoles (top row) with the results of MD with 1\% anti-parallel defect dipoles (bottom row).
 (a) and (b) are ``Field on" cases, while (c), (d), (e) and (f) are ``Field off" cases.
 For the ``Field on" case, the ECE double-peak behavior is present in MC and MD simultaneously.
 For the ``Field off" case both in MC and MD, a single peak is observed under a low field ((c) and (d)), and a shape with one hump and one cusp is demonstrated under a high field ((e) and (f)).
 Therefore, a qualitative agreement between the MC and MD simulations exists.
 }
 \label{fig:mcmd}
\end{figure*}

The coupling to the strain is not considered directly in the MC simulations. 
In order to justify this simplification and the validity of the results obtained, MD simulations in 3D have been performed deploying the feram code~\cite{2008_Nishimatsu} based on an effective
Hamiltonian.
This Hamiltonian includes the self energy due to the ferroelectric soft mode, the long-range dipole-dipole interaction energy, the elastic energy, and the energy due to the coupling between the local soft modes and the strain.
Details of this Hamiltonian were explained in Refs.~\onlinecite{1994_King-Smith,1995_Zhong,2008_Nishimatsu}.
It should be noted that only three degrees of freedom are explicitly taken into account in the MD simulations (the strain is optimized for each step but not used as degree of freedom). Thus, the obtained temperature change overestimates the adiabatic temperature change. 
The leading error can be corrected by a rescaling of the degrees of freedom with 3/15 (actual number of degrees of freedom/degrees of freedom of BaTiO$_3$).
The set of parameters for the effective Hamiltonian for BaTiO$_3$ have been obtained by density functional theory simulation at $T=0$\,K and are
listed in Ref.~\onlinecite{2010_Nishimatsu}. 

Within the MD simulations, periodic boundary conditions and a cell size of 38.4\,nm$\times$38.4\,nm$\times$38.4\,nm have been used.
For simulations in the canonical ensemble ($NPT$), the Nos{\'e}-Poincar\'e thermostat is applied.\cite{1999_Bond} 
Defects are modeled by freezing the local soft mode on random positions to a certain value.
In the following 1\% polar defects anti-parallel to the external field are frozen to mimic a soft-mode amplitude of 0.2\,{\AA} (approximately 0.5\,C\,m$^{-2}$).

The mean distance between defect dipoles in the 2D MC simulations is larger than that in the 3D MD for the same concentration. Namely, in 3D more neighboring defect dipoles appear in the vicinity of non-defect dipoles than in 2D, which leads to a stronger influence of the defects. 
Hence, the defect concentration, as well as the strength of the defect dipoles, should be set smaller in the MD case to achieve similar effects of defects.
In addition, the different energy contributions used in MC and MD, as well as the different parameterizations, result in a quantitative difference in the coupling between the external field and the dipoles.
Due to all these reasons, the strength of the electric field has to be different as well. 
Nevertheless, qualitative agreements can be reached, and even the quantitative agreement of $\Delta T / E $ is observed (see following).

For comparison, MC and MD simulations are carried out for three different cases. The first is a ``Field on" case with a strong field. The loading history of a "Field on" case has been explained in Subsec.~\ref{subsec:negative}. The other two are ``Field off" cases with a weak and strong field, respectively. The so-called ``Field off" case involves the following loading history. Firstly, the system is equilibrated at constant temperature under a fixed external electric field that is anti-parallel to the defect dipoles. 
Secondly, in MC the field is removed instantaneously, while in MD the field is ramped down with a rate of 0.001\,kV/cm/fs.
Simultaneously, the microcanonical ensemble is utilized at a constant energy, using the Multi-demon method in MC~\cite{2012_Ponomareva} and the leapfrog method in MD.
Hereby, the final state at the end of the stage with constant-temperature is used as the
initial state for the microcanoncial ensemble. 
Thirdly, after the system reaches equilibration, the thermal energy is obtained so that the temperature change can be evaluated.
In MD the kinetic energy is monitored after 500,000 further equilibration steps.
By this slow ramping and long equilibration the system can overcome potential energy barriers due to domain switching and is not stuck in a meta-stable state.
Finally the temperature can be calculated by sampling the kinetic energy spatially and historically in both MC~\cite{2015_Ma} and MD simulations.
A time step of $\Delta t = 1$\,fs in MD is used in both ensembles.
It is worth mentioning again that the results produced by the slow ramping are only slightly different from those by the instantaneous ramping.~\cite{2016_Marathe}
Therefore, the results are comparable between MC and MD even though different loading histories are utilized.

It should be noted that the phase transition from the paraelectric to ferroelectric state in MC lies around 400\,K.
However, in the MD simulations the ferroelectric-paraelectric phase transition temperature is systematically underestimated (300\,K) compared to the 400\,K found in experiment.~\cite{2008_Nishimatsu} 
Nevertheless, the aim is to compare the results of MC and MD qualitatively rather than quantitatively.

For the ``Field on" case under a high field, both the MC and MD simulations demonstrate a double-peak ECE behavior, as it is seen in Fig.~\ref{fig:mcmd}(a) and (b). The reason for the double-peak has been explained in the discussion of Fig.~\ref{fig:group}(d). Note that in the MD simulations, the sample is not prepoled. Since the internal fields induced by the defect dipoles are fairly strong, after 500,000 equilibration steps the domain configurations before and after poling have small differences. 

For the ``Field off" case under a weak field, both the ECE curves obtained by MC and MD simulations show a single peak as shown in Fig.~\ref{fig:mcmd}(c) and (d). 
In the MC results the peak appears at 56.4\,kV\,mm$^{-1}$, and in MD at 8.5\,kV\,mm$^{-1}$. The underlying physics for this peak behavior has been explained in the discussion of Fig.~\ref{fig:group}(c) and (d).

For the ``Field off" case under a high field, the curve experiences a hump and a cusp, which is captured both in the MC and MD simulations (see Fig.~\ref{fig:mcmd}(e) and (f)).
The explanation of this peculiar ECE behavior can be referred again to the entropy change in this process.
Below the phase transition temperature, the defect dipoles dominate the domain patterns, and the configuration is a single domain when no external fields is applied. At very low temperature (e.g. 120\,K), the external electric field induces disorder. After removing the fields the configurational entropy thus decreases slightly, which results in a small increase of the vibrational entropy, i.e. of the temperature.
With increasing the initial temperature, i.e. the thermal fluctuations, it becomes much easier to trigger disorder under electric fields. Thus the change of the domain patterns after removing the external fields are more prominent, i.e. the decrease of the configurational entropy is higher during the removal of the fields.
Hence, the increase of the vibrational entropy and of the temperature is more significant.
This phenomenon corresponds to the temperature below 240\,K in MC and below 180\,K in MD.
When the initial temperature is elevated further, due to the higher thermal fluctuations the domain patterns are more disordered in the sample both with and without fields.
It signifies that the difference between the configurational entropy with and without fields becomes smaller.
Therefore, the positive ECE weakens when the initial temperature is increased from 240\,K to 400\,K in MC and from 180\,K to 280\,K in MD.
In other words, the positive ECE is largest at 240\,K in MC and at 180\,K in MD.
When the temperature is high enough, the thermal fluctuation makes it fairly easy to reach an ordered state under assistance of the applied field. Then the disorder increases if the external field is removed, which leads to an increase of the configurational entropy and a decrease of the temperature. 
This phenomenon is common in both the MC and MD simulations above the corresponding para- to ferroelectric transition temperature.
The upcoming cusp appears in MC at 440\,K and in MD at 400\,K.
The existence of the cusp is due to the fact that the phase transition point is shifted to higher temperature in the ``Field off" case. This is similar to the shift of the transition temperature in presence of parallel defect dipoles for the ``Field on" case in Fig.~\ref{fig:group}(a).

%
\begin{figure*}[!htp]
 \centering
 \centerline{\includegraphics[width=13.5cm]{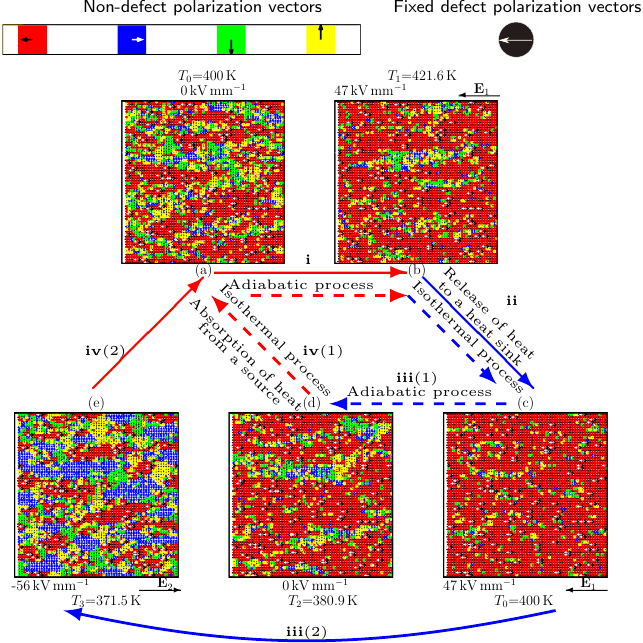}}
 \caption{Proposed modified ECE cycle to enhance the resultant $\Delta T$ of a sample containing defect dipoles. The conventional cycle involves a-b-c-d-a (dashed arrows), and the new proposed one includes a-b-c-e-a (solid arrows). Under adiabatic condition a field is applied rapidly parallel to the defect dipoles, taking the initial state (a) at $T_0$ to a state (b) with lower configurational entropy at $T_1$. By transferring heat to a heat sink, the state (b) transforms to the state (c) with further lower configurational entropy at $T_0$. Removing the field rapidly under adiabatic condition, from state (c) the system is cooled to state (d) at $T_2$. By comparison, applying a field anti-parallel to the defect dipoles the system is cooling further from state (c) to state (e) ($T_3 < T_2$). It should be noted that the magnitude of the external field should not be greater than the magnitude of the average internal field induced by the defect dipoles. Finally, by absorbing heat from a load, the system warms up 
to the state (a).
 The relation of the involved temperatures can be summarized as $ T_3 < T_2 < T_0 < T_1 $.
 It can seen that through introduction of the defect dipoles, an electrocaloric cooling enhancement is achieved.}

 \label{fig:ececycle}
\end{figure*}

In summary, the MC and MD simulations show good qualitative agreements in the three different loading cases.

\subsection{Defect engineering for the ECE} \label{subsec:defect}
Some researchers~\cite{2012_Ponomareva,2013_Axelsson} have revealed a negative ECE induced by noncolinear-field-driven phase transition, and pointed out that this phenomenon can be utilized to enhance the ECE. Similar to that concept, a parallel or anti-parallel electric field is applied with respect to the direction of the defect dipoles. 
Nevertheless, in this work the main focus of interest is on the influence of defect dipoles, as shown in the previous subsections. Based on the gained knowledge the strategies to tailor the ECE by defect engineering can be considered. It is of special importance since defects are unavoidable in real material systems. 

Fig.~\ref{fig:group}(a) indicates that at high temperature the samples with defect dipoles can generate a higher ECE than that without defects under a parallel external field. For example, under an electric field of 47.0\,kV\,mm$^{-1}$, the temperature change at 460\,K for the sample with 2\% parallel defect dipoles is 12.1\,K while only 7.2\,K for the sample without defects. In other words, the parallel defect dipoles have a potential for high temperature applications.

More importantly, the design of a novel electrocaloric cycle can be stimulated by utilizing the positive-negative ECE transition. As an example, we propose in the following a modified Carnot-like Cycle with a reversed external field, which provides the possibility to improve the cooling efficiency of the electrocaloric device. 
A cycle with 3\% defect dipoles at an initial temperature $T_0=400$\,K is illustrated in Fig.~\ref{fig:ececycle}. In the classical concept the ECE cycle~\cite{2011_Scott} includes four states ``a-b-c-d-a" (see dashed arrows), while the new concept contains four states 'a-b-c-e-a' (see solid arrows). Compared with the state (d), the new state (e) involves a reversed electric field, which should induce a further temperature drop, due to the negative ECE indicated in Fig.~\ref{fig:group}(d). 
 
The red and blue arrows represent the heating and cooling processes, respectively.
Process $\mathbf {\romannumeral 1}$ is an adiabatic process from state (a) with $T_0=400$\,K to (b) with $T_1=421.6$\,K, which decreases the configurational entropy, and increases the temperature by application of fields $\mathbf{E_1}$.
Hereby $\mathbf{E_1}$ is parallel to the defect dipoles.
Process $\mathbf {\romannumeral2}$ is an isothermal process from stage (b) to (c) with $T_0=400$\,K, in which the temperature decreases by releasing the heat to a heat sink.
Process $\mathbf {\romannumeral3}$(1) from state (c) to (d) with $T_2=380.9$\,K and $\mathbf {\romannumeral3}$(2) from state (c) to (e) with $T_3=371.5$\,K are both adiabatic processes.
In process $\mathbf {\romannumeral3}$(1) the configurational entropy increases by removal of a field while in process $\mathbf {\romannumeral3}$(2) by application of a reversed field $\mathbf{E_2}$.
Namely, the direction of $\mathbf{E_2}$ is opposite to $\mathbf{E_1}$.
By application of this reversed field the vibrational entropy decreases more than in process $\mathbf {\romannumeral3}$(1).
The temperature drop in $\mathbf {\romannumeral3}$(2) is 28.5\,K, which is higher than 19.1\,K in $\mathbf {\romannumeral3}$(1).
Process $\mathbf {\romannumeral4}$(1) from state (d) to (a) and $\mathbf {\romannumeral4}$(2) from state (e) to (a) are both isothermal processes, in which the temperature increases though adsorption of heat from a heat source.
The example shows that the incorporation of defect dipoles and application of a reversed field can enhance the cooling effect by around 50\%.

\section{Conclusion} \label{sec:conclusion}

In summary, our MC simulation results reveal that the ECE can be influenced heavily by the presence of defect dipoles due to acceptor-vacancy associates, which agrees qualitatively with MD results.
The incorporation of the parallel defects provides a application potential for high temperature by increasing the achievable temperature change.
The anti-parallel dipoles act as ``memory elements" which reduce the configurational space in the prepoled sample and thus allow to turn the positive ECE into a negative ECE. Under certain fields both effects can be combined, which offers possibilities to increase the cooling efficiencies of electrocaloric devices, and indicates potential applications of ferroelectric materials with defect dipoles for the ECE.

\begin{acknowledgments}
The funding of Deutsche Forschungsgemeinschaft
(DFG) SPP 1599 B3 (XU 121/1-2, AL 578/16-2) and A11/B2 (GR 4792/1-2) is gratefully acknowledged. 
Moreover, computing time was granted on the Lichtenberg-High Performance Computer at TU Darmstadt.
Additional computational resources have been provided by the Center for Computational Science and Simulation (CCSS), University of Duisburg Essen.
The authors thank Dr. Min Yi in TU Darmstadt, Yinan Zuo in RWTH Aachen and Prof. Inna Ponomareva in University of South Florida for insightful discussions.
\end{acknowledgments}

%

\bibliographystyle{apsrev4-1.bst} 

\end{document}